# On Engineering and Emergence

Jochen Fromm
Distributed Systems Group, Kassel University
EECS Department for Electrical Engineering and Computer Science
Wilhelmshöher Allee 73, D-34121 Kassel, Germany

fromm@vs.uni-kassel.de

## ABSTRACT

The engineering and design of self-organizing systems with emergent properties is a long-standing problem in the field of complex and distributed systems, for example in the engineering of self-organizing Multi-Agent Systems. We examine the question if a general solution to the MML problem of AOSE and ESOA exists and if a formal approach is possible. The problem of combining engineering with emergence – to find a simple rule for a complex pattern – equals the problem of science in general. Therefore the answers are similar, and the scientific method is the general solution to the problem of engineering complex systems.

## KEYWORDS

Distributed system, Self-Organization, Science, Engineering, Theory, Analysis, Design, Model, Cellular Automata (CA), Multi-agent system (MAS), Emergence, Micro-Macro-Link (MML), Agent-Oriented Software Engineering (AOSE), and Engineering Self-Organizing Applications (ESOA)

## 1. INTRODUCTION

Complex systems with emergent properties are fascinating. And they are very fundamental. If we consider the fact that everything is made out of elementary particles – quarks, electrons, photons, etc. – then every macroscopic object and law we know is the result of some emergence process. According to J.M. Ottino [1], "the hallmarks of complex systems are adaptation, self-organization and emergence – no one designed the web or the metabolic processes within a cell". Emergence is in fact a basic property of many distributed and complex systems [2,3].

A desirable feature of emergent properties on the macroscopic level is insensitivity to microscopic processes and independence of individual elements or components. They are useful to achieve robustness, scalability, autonomy and self-* properties (self-healing, self-optimization, self-configuration,…) in systems. Therefore a method to engineer systems with emergent properties would be very useful, especially for Cellular Automata (CA) and Multi-Agent System (MAS) which are often only used to describe and simulate complex systems [30,31].

Yet there is no standard engineering method for systems with emergent properties. They seem to resist any attempt to design and engineer them systematically. As Jennings, Sycara and Wooldridge say in their "Roadmap of Agent Research and Development" [24], "one must use a laborious process of experimentation, trial and error" to engineer such MAS. Does the combination of engineering and emergence make sense – can we design a predictable system with unpredictable properties – or is it a contradiction in itself? Is the engineering of self-organizing systems with emergent properties [1] possible at all?

It is difficult to engineer a system which creates and makes its own laws. It seems hardly possible to organize a system that organizes itself. We know that nature has created such self-organizing systems (for example ant and termite colonies, etc.) through evolution, and we can create similarly self-organizing systems by evolutionary algorithms. Thus one solution is the use evolutionary and genetic algorithms.

In this article we examine the question if an "Intelligent Design" of self-organizing systems is possible in general, too. The problem in the engineering of many complex systems is emergence, and the central question of emergence is how you can use simple local rules to generate specific higher levels of organization. Since this is unclear in general, the phenomenon of "emergence" is at odds with any attempt to engineer and design a system.

Yet the situation is not totally hopeless, because the central question mentioned above is also the core problem of science in general: to find simple rules for complex patterns, to uncover simple laws behind complex behaviors, and to discover simple theories for complex phenomena. John H. Holland says [3] "The hallmark of emergence is this sense of much coming from little. This feature also makes emergence a mysterious, almost paradoxical, phenomenon". As Einstein said, mysterious things are also the source of science. It is the goal of science to explain and uncover mysterious things. Something is mysterious, if we observe an effect without knowing the cause, if the knowledge of the causal connections is incomplete. This is typical for emergent properties, too.

Thus the scientific method may offer a preliminary answer to the problem of engineering complex systems, i.e. to the combination of engineering and emergence. Although this may sound like a contradiction to a scientist, the answer is probably an *intelligent design* based on the classic *scientific method*, as we will see in detail later. We will draw parallels to fractal geometry and other fields to support this thesis.

---

[1] In this paper, the terms self-organization and emergence are interchangeable and we do not make a distinction between them, although the former refers more to a dynamic process across the boundary between system and environment, and the latter more to a process across the boundary between microscopic and macroscopic regions. In practice both occur often together and are characteristic for many complex systems in nature which are the result of evolution and not the product of goal-directed engineering or planned design.

The rest of the article is organized as follows: section 2 examines the various problems related to the combination of the two basic concepts engineering and emergence, section 3 proposes a general solution and four specific examples for an approach, and section 4 summarizes the results in a short conclusion.

## 2. PROBLEMS

### 2.1 Understanding of the Principles

The ultimate problem related to fields engineering and emergence is of course to understand the principles of emergence. If we understand the principles well enough, then even the engineering of systems with emergent properties should be possible. How far we can understand the principles is still an open question, for example what are the limits of self-organization and emergence? How complex can a system become without evolution or evolutionary processes, only by repeated interactions of simple agents or repeated iterations of simple rules?

Since emergence is basically a bottom-up process, probably we have to consider the problem from the bottom-up view: How can we generate complex global behavior from simple local actions? How can we *design* local behavior so that a certain global behavior emerges? How can we *engineer* self-organizing systems? In order to this, we need to understand, control, and generate emergent behavior. Robert Axelrod formulated it in this way: "how do you use simple local rules to generate higher levels of organization from elementary actors?" [4].

A solid understanding of this problem would simplify science and engineering, since complex systems are their basic subject. In science we want to understand complex systems, in engineering we want to construct them, in the former we want to explain complexity through a simple rule, equation or formula, in the latter we want to hide complexity behind a simple interface.

Pattie Maes mentioned the question of understanding in her classic paper from 1994 in the conclusion [5]: "We need a better understanding of the underlying principles. In particular, it is important to understand the mechanisms and limitations of emergent behavior. How can a globally desired structure or functionality be designed on the basis of interactions between many simple modules? What are the conditions and limitations under which the emergent structures are stable, and so on". More than ten years later, these questions are still a matter of research. Scientists are dealing with exactly the same "local-to-global" or "micro-to-macro" problem - see for example [8]. A solution is certainly not easy.

### 2.2 Degree of Predictability

It is a problematic and interesting question to what degree a complex system created by a CA or MAS is predictable at all. In the mathematical Chaos Theory deterministic rules can lead to non-deterministic, random and chaotic behavior. Systems with very simple rules can produce complex strange attractors. The fascinating thing in Chaos Theory is the complexity of the strange attractors, which arises from order in disorder or determinism in randomness or chaos. Small variations in initial conditions do lead to large variations in the long run, but the trajectories stay always near the attractor. Only the motion *on the attractor* exhibits sensitive dependence on initial conditions. We can predict the motion or the path of the trajectories for very small time scales (through corresponding differential equations), and the average position for large time scales (somewhere on the attractor), but we can not predict the dynamical state for intermediate times scales (where exactly on the attractor is not known).

In other words the future may be different in details - the position on the attractor - but certain in general aspects. This predictability in unpredictability or regularity in irregularity is typical for complex systems. They are predictable and regular only at certain scales, but unpredictable and irregular at other scales. The weather for example can be predicted well for the next day, or roughly for each of the four seasons, but not very well on intermediate scales.

Probably we cannot define and predict exactly what a complex system looks like at each time step on the macroscopic level (due to combinatorial explosion and other effects). But we can define roughly what the system does: what roles the agents occupy, what type of attractors and emergent phenomena appear – but not when and where exactly. We can also predict how certain types evolve and change roughly – but not in every detail, only on certain time scales. As Jonathan Rauch argued [9], we will probably not be able to foresee the future of each possible MAS in every detail, but we might learn to anticipate the kinds of events that lie ahead.

The engineering of self-organizing systems with emergent properties does only make sense if we can predict the amount of unpredictability and learn to understand it, if we know what kind of emergent properties can appear (but not necessarily when or where exactly).

### 2.3 Autonomy and Purpose

Engineering is associated with a process of precise composition to achieve a predictable purpose and function. It is all about "assembling pieces that work in specific ways" [1]. Engineering means imposed purpose and function, contrary to autonomy which implies unclear or changing purpose and context-dependent or emerging function.

In engineering parts with a certain function are composed for a certain purpose. In complex systems which are described by CA or MAS the purpose, function and role of each part is not clear. A cell in the Game of Life can be a part of a Blinker for a long time and in the next moment in can remain silent and frozen, i.e. it stays always on or off. This can happen for instance if the Blinker collides with a Glider or a Spaceship. Therefore it is hard to define a fixed purpose or a function for a specific cell.

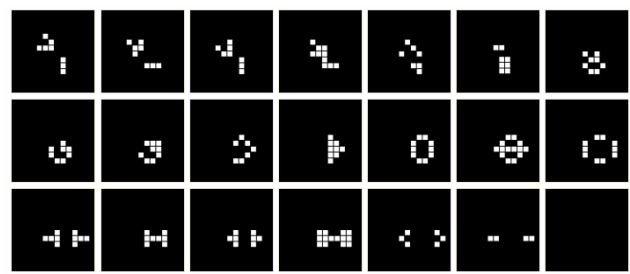

**Fig. 1 Collision of a Glider and a Blinker**

In agent-based systems, the situation is similar. An ant in an ant colony can play different roles; it can be a soldier, nest builder, explorer, transporter or path follower, depending on the context and situation. The role, task and function of the agents, elements and components can change, unpredictable or even paradox phenomena can emerge, and the purpose, objective and goal of the overall system can not be obtained as a simple sum from the individual components. Depending on the goals and roles of the agent, it has various functions in different contexts, and it may act against the interests of the system or even start to destroy it.

This basic conflict between the interests of the elements or agents and the interests of the system is the AOSE problem: can we unite autonomy and engineering? The more autonomous an agent is, the greater is the freedom to act independently, and the less is a certain function or guaranteed purpose in respect of a larger system.

## 2.4 Imposed- and Self-Organization

The name AOSE is the name ESOA read backwards, but the one is not the solution of the other. They are both words for problems, not for solutions. Engineering means imposed organization, which is obviously opposed to self-organization. This is the ESOA problem: can we combine self-organization and engineering? The difficulties in the design and engineering of self-organizing MAS are certainly not accidental. It seems to be an intrinsic and inherent problem related to MAS and autonomous agents in general, because engineering and autonomy interfere with each other, and other factors like evolution and "emergence" make the control of such systems difficult. Edmonds says [7]: "Engineering and Self-Organization do not sit well with each other. The extent to which a system is engineered will constrain (as well as enable) what kind of self-organisation can occur."

## 2.5 Top-Down or Bottom-Up

In traditional object-oriented software systems we have well established methodologies to design and engineer systems. For complex systems that are simulated or created by CA or MAS the situation is different, because the question of the Micro-Macro-Link (MML) is much more complicated here. In these systems, the usual top-down Macro-to-Micro direction - known as "design methodology" or "analysis" - is not enough. The unpredictable bottom-up Micro-to-Macro direction in this type of complex systems makes any pure top-down attempt useless. A bottom-up approach alone is not feasible either. Even if we consider simple CA or MAS, the number of combinations and configurations grows exponentially with the number of states, elements, and rules.

In traditional object-oriented software systems, there is no Micro-Macro-Link (MML) problem. In agent-oriented systems this is different. On the conceptual level, a collection of software objects may be viewed as a machine, where each part has a well-defined function at an unchanging place, whereas a collection of agents is more like a flexible society or dynamic community. Organizations are the natural way to organize a society. Yet how can we infer the right organizational structures from the global requirements, if the system is allowed to organize itself?

Should we use a top-down or bottom-up approach? As we have seen and argued in the introduction, neither is sufficient alone. How can we combine both, an imposed top-down organization from the outside, and an emerging bottom-up self-organization from the inside?

## 2.6 Formal or Experimental

There are two obvious solutions to build a self-organizing system that meets the requirements and objectives: first the imitation of natural systems, for instance in form of biologically and sociologically inspired system, or second manual trial-and-error. The first method can only be applied to transfer existing solutions, the second is not systematic, as Edmonds says [7], "we are to do better than trial and error…we will need to develop explicit hypotheses about our systems and these can only become something we rely on via replicated experiment."

Experiments, simulation and modeling are essential in order to cope with these problems. Just as we can simulate natural systems to understand and model them (e.g. biological or social systems) we can simulate artificial systems to understand and engineer them. Edmonds and Bryson [6,7] have recommended a methodology based on an experimental method – to make testable hypotheses about the behavior of the system that must be verified by experiments, because formal methods are limited and there is no effective means of finding a system that satisfies a given formal specification.

It is clear that an experimental approach alone is not enough, we need also a formal element to focus on the requirements and global objectives. Otherwise the number of possible systems, combinations and configurations is simply too large. How can we combine both in a serious and sensible way, an analytic approach and an experimental method?

However, existing AOSE formal methodologies lack this experimental method. They also usually do not consider emergence and self-organization, since they require the definition of static organizations, roles and organizational structures, contrary to self-organizing systems with emergent properties in nature, which grow, evolve and even organize themselves.

## 2.7 Existing Methodologies

The large number of different Agent-Oriented Software Engineering (AOSE) Methodologies indicates that there is a big need for such a methodology, and yet at the same time that there exists a major obstacle: engineering (in form of software applications) and autonomy (in form of autonomous agents) don't seem to fit well together. A large number of methodologies claim to solve this problem. The major players are:

- **ADELFE** (Atelier de Développment de Logiciels à Fonctionnalité Emergente) from Bernon and Gleizes [10]
- **GAIA** $1^{st}$ version from Wooldridge, Jennings & Kinny [11], refined version from Zambonelli, Jennings & Wooldridge [12]
- **Tropos** $1^{st}$ Version from Bresciani, Perini, Giorgini, Giunchiglia, Mylopoulos, [13], refined version from Kolp, Giorgini and Mylopoulos [14]
- **MASSIVE** (Multi Agent Systems Iterative View Engineering) Lind [15]
- **MaSE** (Multiagent Systems Engineering) from Wood and DeLoach [16]
- **Prometheus** from Padgham and Winikoff [17]

ADELFE is one of the few methodologies which claim to support emergent behavior, since "emergent functionality" is already part of the name (Atelier de Développment de Logiciels à Fonctionnalité Emergente) [10]. Yet it remains to be shown that you can really construct interesting new forms of emergent behavior with Adelfe. MASSIVE and Prometheus are nearly the only methodologies which mention the need for an iterative process. Especially MASSIVE emphasizes the importance of "Round-trip Engineering" and "Iterative Enhancement". A stepwise refinement and iteration is probably essential to construct systems with emergent properties.

Carlos Gershenson has proposed "a General Methodology for Designing Self-Organizing Systems" [22] based on the five basic steps representation, modeling, simulation, application, and evaluation. If a general methodology to engineer self-organizing Multi-Agent Systems exists, it will probably contain these five steps and especially the formulation of testable hypotheses, as Edmonds and Bryson said [6,7].

It will also contain the emphasis of "emergence" from ADELFE [10], the "Round-trip Engineering" and "Iterative Enhancement" known from MASSIVE [15], the concepts of roles and organizational structures from GAIA [12], and the highlighting of goals and multi-agent patterns from TROPOS [14]. It can not be a pure top-down method, because these systems have unpredictable properties which can only be detected by simulations and experiments. Instead we should expect some kind of round-trip process based on stepwise iterative enhancements which bridges the micro-macro distinction. Such a method indeed exists, and it is well known, as we will see now.

## 3. APPROACH

*"The aim of science is, on the one hand, a comprehension, as complete as possible, of the connection between the sense experiences in their totality, and, on the other hand, the accomplishment of this aim by the use of a minimum of primary concepts and relations".*    Albert Einstein

### 3.1 Solutions

There is no formal approach or methodology which would be a solution to the problems of "engineering emergence" or "engineering self-organizing systems", but there is a well-known general method based on a stepwise iterative enhancement which bridges the micro-macro distinction – the classic scientific method. Of course the scientific method is well-known and has been taught and applied for centuries. Yet it has not been applied directly to engineering problems. Engineering is about the application of established scientific knowledge. The scientific method is usually found in the natural sciences like physics, chemistry and biology which use experiments to gather evidence and numerical data in order to validate hypotheses and theories.

The point is that general distributed systems – whether they are described and implemented by a Cellular Automata (CA), Multi-Agent System (MAS) or Distributed Algorithm (DA) – can grow so complex that they constitute their own little world with their own macroscopic laws. Andrew Ilachinski has named his book about CA strikingly "*Cellular Automata: A Discrete Universe*" [31]. One has to apply the scientific method directly to these artificial worlds to find out the appropriate laws and models in each case. There is no reason why the scientific method cannot be applied to artificial worlds. Once we know the particular universe and the elements and laws in it, for instance all the blinkers, spaceships and gliders that are possible in a particular system, we can start to use and combine them in order to engineer a new system.

This does not mean that every CA, MAS or DA produces a very complex range of phenomena. Quite the contrary, a simple rule which produces complex results is rather the exception than the norm. From all the 256 elementary one-dimensional CA, only a few are really complex, ∼2−5% depending on your classification and point of view. Wolfram classified them as class IV, Eppstein as a class where both expansion and contraction possible[2]. These rare forms occur like Conway's Game of Life at the 'edge of chaos' between order and randomness and are neither completely random and chaotic nor completely repetitive and uniform. They are of course the most interesting types. Although each CA can be considered as a discrete universe that has a certain dimension and extension, and its own rules, laws and states, only a few are really interesting. Each of them is a small universe or miniature world of its own which has to be investigated by the scientific method.

The same argument can be applied to complex board games like chess, tic-tac-toe, or Go where the number of configurations is very large despite rules which forbid many possibilities [3]. Although the rules are very simple, these games are a world of their own which have fascinated humans for centuries, and sometimes they still intrigue us.

Thus the distinction between science and engineering blurs increasingly for the engineering of complex and self-organizing systems, and both fields seem to converge more and more in this case [25]. In the engineering complex self-organizing systems, the goals of science (to explain and to understand complexity) and engineering (to hide and to master complexity) merge; we construct the complex miniature world we want to understand at the same time. Certainly we cannot master the engineering of a complex system if we do not understand it.

### 3.2 The Scientific Method

There are two simple reasons why science is possible. First, as John D. Barrow says in his book *Impossibility* [27], "Science exists only because there are limits to what Nature permits. The laws of nature and the unchanging constants of nature define the borders that distinguish our universe from a host of other conceivable worlds where all things are possible". The laws governing nature help us to separate the possible from the impossible. Just as in simple board games, not all moves or configurations are legal: "The rules constrain the possibilities" [3].

Second, simple computational rules can like simple mathematical equations generate complex, seemingly unpredictable random-looking behavior. It is perhaps the most fundamental idea that unifies science. Without this fundamental fact, science would be impossible, because we would never be able to explain, to describe or to predict complex phenomena in terms of simple rules or equations. In other words, regularities in nature exist and these regularities can be described by simple laws or rules.

---

[2] See http://www.ics.uci.edu/~eppstein/ca/wolfram.html

In mathematics and physics it is a well-known fact that simple systems or equations result in very complex behavior, especially in "Chaos Theory". The simple equation $z = z^2 + c$ produces the complex Mandelbrot set, Newton's simple formula for Gravitation $F = G\, m_1 m_2 / r^2$ describes all the motions and actions in a complete solar system. Steven Wolfram emphasized and illustrated this fact ad nauseam in his NKS book [30] for complex systems that can be described by CA: "a simple program can produce output that seems irregular and complex". John Holland has noticed this for agent based systems and board games like chess [3]: "a small number of rules or laws can generate systems of surprising complexity".

The problem of "engineering emergence" - to find a simple rule for a complex pattern – equals the problem of science in general: to explain complexity by describing complex natural phenomena with a minimum of primary principles, laws and rules. David Hilbert (1862-1943) said "The art of doing mathematics consists in finding that special case which contains all the germs of generality". The art of science is to find that special theory which describes and explains as many complex phenomena and observations as possible: "One of the principal objects of theoretical research in any department of knowledge is to find the point of view from which the subject appears in its greatest simplicity" (J. W. Gibbs).

---

1. **Observation/Examination**:

   Observe and measure some aspect of the world

2. **Hypothesis/Theory Formulation**:

   Invent or modify a theory, a law or a hypothesis

3. **Prediction/Evaluation**:

   Use the hypothesis to make predictions

4. **Test/Experimentation**:

   Verify and test the predictions by experiments

**Fig. 2 The Scientific Method**

---

Science comes from Latin scientia - knowledge - and refers to a system of acquiring knowledge[3]. The basic unit of knowledge in science is the (mathematical) theory, model or metaphor. The basic unit of knowledge in complex systems represented by agent-based simulations is the (agent) model, a combination of simple local rules and complex global behavior. The task is to find the rule of the 'game', not only in board games and evolutionary game theory where this is literally true.

The difficulty is of course to discover the right theory or model for the phenomena in question - to find the corresponding simple rules for a given system. Not every simple rule gives rise to a complex system, and not every simple rule which does this is appropriate for a given system. The formulation of the hypotheses, laws, principles and theories in science is the big, non-trivial step based on personal experience. This hard step requires a lot of creativity, intuition and curiosity. A theoretical scientist needs to play around with as many original, eccentric and unusual ideas as possible. In evolutionary algorithms, this type of creativity is caused by recombination and mutation. In science it is achieved not only by the ingenious ideas of individual scientists, but also by the interaction among peers and the constant merging and splitting of publications and scientific ideas.

Science is basically a two-way or two-phase activity, divided into theory and practice. Theoretical science or theory is traditionally a top-down examination to explain, predict and describe values which have been measured in experiments (see Fig. 3), whereas experimental science is a bottom-up try to verify theories. Experimental scientists in the applied divisions observe and gather data, whereas theoretical scientists invent and discover theories.

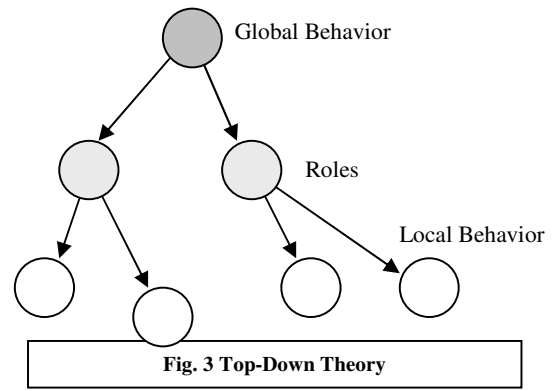

**Fig. 3 Top-Down Theory**

The most important feature of science is possibly the verification of theories, laws, principles and hypotheses through experimental tests. This requires a constant comparison of theory with experiment (see Fig. 4) and implies a continuous round-trip from the concrete (usually macroscopic) world to the abstract (usually microscopic) model and back, resulting in an iterative refinement of the theory or model. Just as a scientific theory is worthless without experimental verification, an agent model for a self-organizing system with emergent properties is worthless without extensive experimental tests and verifications.

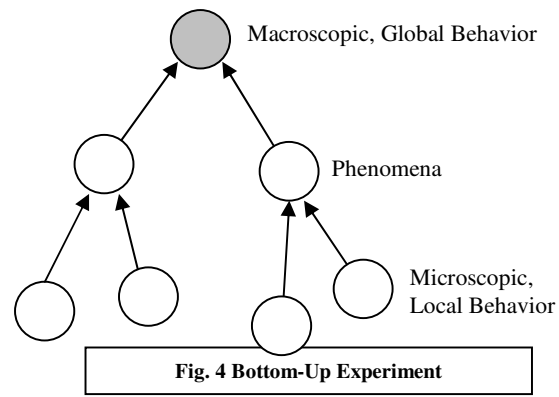

**Fig. 4 Bottom-Up Experiment**

So the good news is: there is a method for the engineering of self-organizing systems with emergent properties. It is the well-known good old *scientific method*, only applied to an artificial world instead of natural world. The bad news is: the approach is not a formal approach or detailed recipe, it does not offer explicit guidance or a complete construction manual, and relies on human experience and creativity. It is hard to standardize creativity and imagination. Contrary to other solutions like evolutionary algorithms, the performance of the approach is certainly difficult to measure: you cannot predict how fast a scientist will solve a problem.

---

[3] See http://en.wikipedia.org/wiki/Science

The only other comparable approaches to engineer complex systems in mathematics and computer science work the same way, for instance the "inverse problem" in fractal geometry, Rapid Prototyping or Agile Software Development [29], or Pattern-Oriented Modeling in Agent Based Systems [26]. And they have similar drawbacks: they require considerable human intervention, are highly non-trivial and difficult to automate. The scientific method is no silver bullet and gives no guarantee that a solution for a specific problem exists: it can be found immediately or not at all.

It is an abstract method. As we will see in the following, you might call the concrete method derived from the general *scientific method* stepwise refinement of educated guesses or …

- the graduate student method
- interactive man-machine method [28]
- iterative two-way approach [18]
- synthetic microanalysis [19]
- goal-directed simulation [23]

## 3.3  The Graduate Student Method

The inverse problem in fractal geometry (to find a simple fractal rule or iterated function system for a given complex image) is very similar to the problem of "engineering emergence" and science in general (to find a simple rule for a complex pattern) and it is so hard that Michael Barnsley's algorithm to solve it was derisively referred to as such a "Graduate Student Algorithm" [4].

> 1. Acquire a graduate student
> 2. Give the student the task…
> 3. …and a room with a computer
> 4. Lock the door
> 5. Wait until the student has engineered the system
> 6. Open the door
>
> **Fig. 6 The graduate student algorithm**

This is a simple method to solve any scientific problem in general. Of course the graduate student algorithm is obviously not a suitable or serious general method, but it comes in fact close to the general solution of the problem: the application of the scientific method. The application of the scientific method can of course be observed optimally in the work of scientists themselves. Let us see how some of the great computer scientists have applied it in order to construct and create a self-organizing system.

---

[4] See http://www.faqs.org/faqs/compression-faq/part2/section-8.html

## 3.4  The Interactive Man-Machine Method

One of the great computer scientists is John von Neumann. John von Neumann's self-reproducing CA is certainly an example for a self-organizing system with emergent properties. His ideas about self-replicating patterns inspired later John Conway to his famous Game of Life. Von Neumann found a self-reproducing system with 29 states, Edgar Codd one with 8 states.

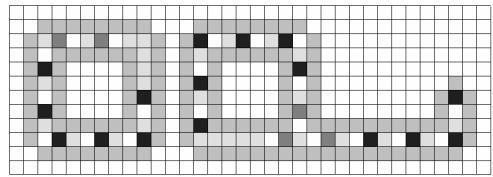

**Fig. 5 Codd's Self-Reproducing CA**

Image from an applet by Martin Heinrich, found at
http://stud4.tuwien.ac.at/~e0125222/codd/, see also
http://en.wikipedia.org/wiki/Codd's_Cellular_Automaton

According to Arthur W. Burks, both used an approach named "the interactive man-machine method" [28]:

> "Codd's cellular automaton system is primarily of interest to us here because it was generated by what we have called 'the interactive man-machine method'. As von Neumann did, Codd chose as sub-goals certain elementary behavioral functions, which he thought he could later synthesize into organs, larger units, and finally universal computers and constructors. He then proceeded to define his transition function piecemeal so as to obtain these behaviors, retreating when a partial definition turned out to have undesirable consequences and either modifying the definition as it had been specified at an earlier stage or seeking alternative behavior (sub-goals) to realize the final goal. But von Neumann proceeded analytically, using only his own reasoning and testing a few cases by hand, whereas Codd used a computing machine to assist him."

The method which Burks describes – the interactive man-machine method – highlights the close interaction between theory and experiment and emphasizes the identification of sub-goals and elementary behavioral functions (roles). It can be summarized like this

> 1. Identify subgoals and elementary behavioral functions
> 2. define preliminary (or modify existing) transition function
> 3. observe simulated behavior, test cases and consequences
> 4. modify definitions
> 5. go back to the beginning → 1.
>
> **Fig. 7 Interactive Man-Machine Method**

# INSERTION: The Inverse Problem in Fractal Geometry

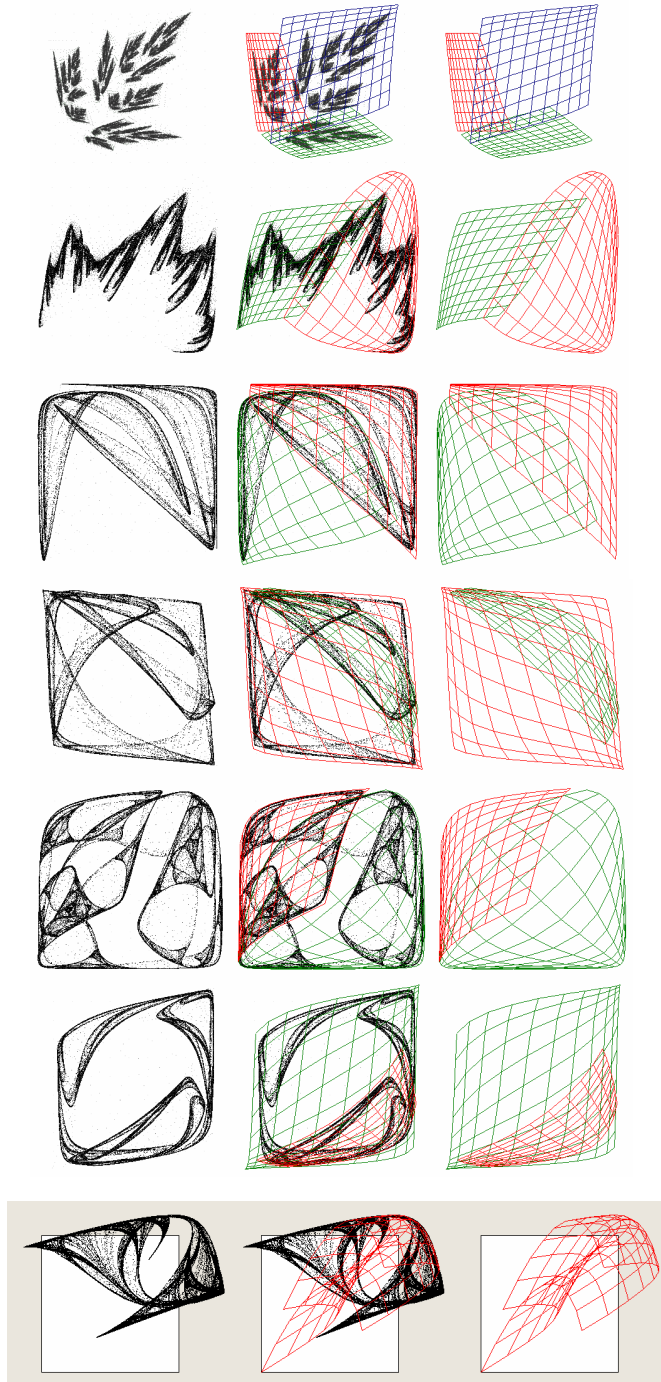

An Iterated Function System (IFS) is a set of functions which define a fractal. Creating the fractal image is easy if you know the functions, but finding the fractal rule or function system for an existing image is a hard task. It is known to be NP-Complete [32], and it is very difficult to solve automatically.

The figure at the top shows a simplified version of Barnsley's fern (only the fourth function for the stem is missing). In each of the figures, on the left we can see the fractal image, on the right the corresponding IFS, and in the middle the combination of both. The problem in fractal geometry is to create the fractal image from the set of simple functions, instructions or rules, i.e. to go from the right to the left. The inverse problem in fractal geometry is to go from the left to the right.

In order to find the rule responsible for the fractal image, it is useful to decompose the image into self-similar parts or copies. The decomposition is a non-trivial part. Splitting the image into parts is not difficult, splitting the image into self-similar parts is a bit more difficult, and splitting the image into the right, minimal number of self-similar parts is very difficult in general.

If the transformations or functions are linear, or nearly linear, then they are only a scaling and rotating of the image as we know from linear algebra. In this case they can be determined by measuring distances and angles.

If the transformation rules of the IFS are non-linear, the inverse problem gets even more difficult to solve, although it is still possible in principle. Measuring a few distances and angles is not enough in the non-linear case, as we can see in the last example at the bottom.

An easy solution for the inverse problem of fractal geometry in general would be a powerful tool to encode and compress all kind of images. For self-similar fractals they offer an unbeatable degree of compression. Unfortunately, many natural images are not perfectly self-similar. Fractal compression has no general advantage compared to other compression algorithms for natural images. Quite the contrary, for many natural images it is much slower - especially the time consuming encoding process - and the degree of compression is not significantly higher.

## 3.5 Iterative Two-Way Approach

After this closer look at fractal geometry and how John von Neumann and Edgar Codd constructed self-reproducing automata, we take a short look on the solution of the micro-macro problem.

As Conte and Castelfranchi have argued, [18] the micro-macro link (MML) problem in general probably needs a two-way or two-phase approach to find the necessary micro-macro connections, including a bottom-up and a top-down process. The way up corresponds to experimental science which means synthesis, simulation or experiments, and determines how individual actions are combined and aggregated to collective behavior. The way down corresponds to theory and means analysis, creation of testable hypotheses, or translation of requirements, and defines how collective forces influence and constrain individual actions.

We can only generate complex self-organizing systems with emergent properties in a goal-directed, straightforward way if we look at the microscopic level and the macroscopic level (for local *and* global patterns, properties and behaviors), examine causal dependencies across different scale and levels, and if we consider the congregation and composition of elements as well as their possible interactions and relations. A complex system can only be understood in terms of its parts *and* the interactions between them, if we consider static *and* dynamic aspects.

In other words we need a combination of top-down and bottom-up approach, which considers all aspects: static parts and dynamic interactions between them, together with the macroscopic states of the system and the microscopic states of the constituents.

An iterative step-by-step cycle based on the scientific method would look like this:

---

1. Define the function or purpose of the system
2. Gather information, review existing solutions
3. *Top-Down*: form model, identify parameters
4. Plan experiment, predict result
5. *Bottom-Up*: do experiment and collect data
6. Analyze and interpret data
7. Draw conclusions → goto 3

**Fig. 8 Iterative Two-Way Approach**

---

In a typical iteration, you start from the "top" and work your way down to the micro-level, constructing agent roles and interaction rules in just the way necessary to generate the behavior observed on "top". If you have arrived at the "bottom", you need to run comprehensive simulations and experiments to verify if the system meets the expectations.

This procedure can be iterated by stepwise refinement of agents and their interactions, which should include necessary changes in the environment, until the desired function is achieved. In the next round, you start start again from the global structure or macroscopic pattern, and try to refine the possible underlying micro-states and micro-mechanisms.

A look to the inverse problem of fractal geometry is maybe helpful here. It is relatively easy for "pure" fractals, and hard for natural images which are not self-similar, because one has to identify possible decompositions - especially self-similar regions and scaled copies of the image itself - to find suitable transformations. In order to create a self-organizing system it is therefore useful to look for self-similar regions, for example regions where a group of agents should act like a single agent.

## 3.6 Synthetic Microanalysis

One example for a two-way approach is the informal method from Sunny Y. Auyang named "Synthetic Microanalysis" (SMA) which claims to combine synthesis and analysis, composition and decomposition, a bottom-up and a top-down view, and finally micro- and macro-descriptions [19]. The general concept is a "bottom-up deduction guided by a top-down view", see Fig. 8. One has to delineate groups of "microstates" according to causal related macroscopic criteria. In other words you try [19] "to cast out the net of macro-concepts to fish for the micro-information relevant to the explanation of macro-phenomena" (p.56). "If you make a round trip from the whole to its parts and back" [19], you can use the desired global macroscopic phenomena to design suitable local properties and interactions.

This two-way approach is a generalization of the "experimental method" proposed by Edmonds and Bryson [6,7]. In the top-down phase you have to create testable hypotheses, which have to be verified in the experimental bottom-up phase. The bottom-up approach alone is successful only for small and simple systems like 1-dim Cellular Automata, where you can enumerate all possible systems. For large systems the number of configurations grows so large (or even "explodes") that the goal gets lost and the thicket of microscopic details becomes impenetrable. To quote Auyang [19] again: "blind deduction from constituent laws can never bulldoze its way through the jungle of complexity generated by large-scale composition" (p.6).

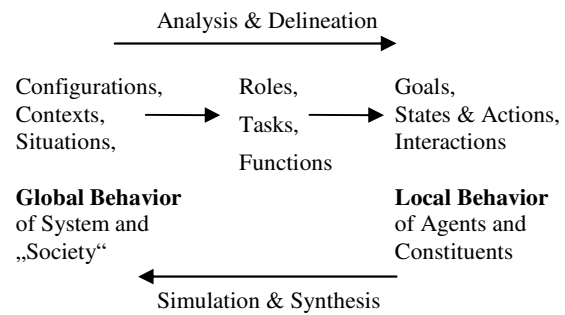

**Fig. 9 Synthetic Microanalysis**

The macroscopic view is useful and necessary to delineate possible configurations, to identify composite subsystems on medium and large scales, to set goals for microscopic simulations and finally to prevent scientists "from losing sight of desired macroscopic phenomena when they are immersed in analytic details" [19].

One round trip from the whole to its parts and back is probably not enough to generate complex self-organizing systems with emergent phenomena. You will certainly need some iterations and a number of stepwise refinements until the method converges to a suitable solution.

It is important to identify and refine before each iteration suitable subsystems, basic compounds and essential phenomena on the macroscopic level, which are big and frequent enough to be typical or characteristic of the system, but small and regular enough to be explained well by a set of microscopic processes. Many macroscopic descriptions are only an approximation, idealization and simplification of real processes.

In the first top-down phase towards the bottom level, we must find the significant, relevant and salient properties, events and interactions. We seek the concrete, precise and deterministic realization of abstract concepts. Many microscopic details are insignificant, irrelevant and inconsequential to macroscopic phenomena. In the second bottom-up phase towards the top level, one has to compare the results of the synthesis and simulation which the desired global structure. The two phases are asymmetric, the first phase needs more manual work and "analytic" considerations, the second requires more computational work and "numeric" simulations.

One would roughly proceed in two phases while trying to determine possible states, roles and role transitions:

**Phase 1. Theory - Analysis and Delineation**

Starting from requirements and global objectives, what macroscopic and microscopic patterns, configurations, situations and contexts are possible in principle? From the answers you can try to delineate what roles, behaviors, local states and local interactions are roughly possible or necessary:

a) What roles and local behaviors are possible? Try to determine and deduce local behavior from global behavior, identify possible roles and role *transitions*.

b) What states are possible? Determine and define local properties from global properties.

c) What kind of local communication and coordination mechanisms are possible? Determine tolerable conflicts and inconsistencies.

**Phase 2. Experiment - Synthesis and Simulation**

Is the desired global behavior achievable with the set of roles and role transitions? The second phase consists of comprehensive simulations and experiments.

Since emergent properties are possible, simulation is the only major way up from the bottom to the top. As Giovanna Di Marzo Serugendo says "the verification task turns out to be an arduous exercise, if not realized through simulation" [20].

Sometimes the term "emergence" itself is even defined through simulation, for instance in the following way: a macrostate is weakly emergent if it can be derived from microstates and microdynamics but only by simulation [21].

The way up is simpler than the way down and requires mainly simulations. Since these simulations can be quite time consuming, it can be slower than the top-down process. In mathematical calculus, the situation is quite similar: many integrals can only be solved and determined numerical by numeric calculations, whereas differentiation is much easier and requires often only sophisticated analysis and analytic techniques.

The principles and processes of SMA and genetic algorithms (GA) are quite similar, see Fig. 10 for a comparison. Both require the use of simulation, experimentation and selection. In the case of evolutionary algorithms without "humans in the loop", the fitness evaluation is done automatically by fitness functions, in the case of synthetic microanalysis with "humans in the loop" it is done by the human engineer.

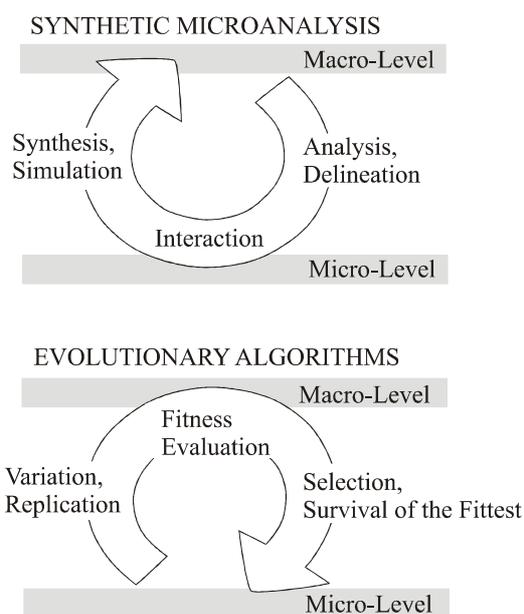

**Fig. 10 Synthetic Microanalysis vs. Genetic Algorithms**

The advantage of *Synthetic Microanalysis* (SMA) is that we are able to understand the solution. The drawback: it is an *informal* method without clear rules for each step, and it still requires a human-in-the-loop. It is based on human intelligence, creativity and experience, and needs - if it works at all - constant manual intervention, observation and consideration.

The advantage of *Genetic Algorithms* (GA) is that they do not require a human-in-the-loop. The drawback is that we are sometimes not able to understand the result. It is often hard to understand why the result is optimal (and none of the other solutions) and how it works exactly.

Yet this comparison is misleading. Unfortunately SMA is not a formal algorithm or method for the engineering of self-organizing systems. It is doubtful if it does deserve to be called a 'method' at all, because it is only an informal description of an undetermined process: the application of the scientific method by a single scientist or engineer. It relies on human experience and creativity, requires considerable human intervention, and gives no guarantee that a solution for a specific problem exists. The solution can be found immediately or not at all.

## 3.7 Goal-Directed Simulations

The engineer who applies the scientific method to create a self-organizing Multi-Agent System (MAS) is in no way different from the social scientist, who applies the scientific method to MAS as well. Hence it is not surprising that the basic loop in all above methods and approaches can also be found in any Multi-Agent Based Simulation (MABS), esp. in the social sciences where a Multi-Agent System (MAS) is used to model systems with multiple social actors.

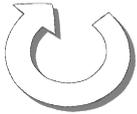

Bruce Edmonds describes the basic MABS sequence like this [23]: (1) the MAS is designed to incorporate the relevant aspects (2) the MAS is run (3) the resulting process is analyzed, and conclusions about the behavior displayed in the MAS "run back in terms of that system". In other words, interpretations and conclusions of the simulation are used to refine the model. Such a closed sequence is quite similar to one cycle or round-trip in the approaches mentioned above.

Social systems are often very messy. As Scott Moss noticed [24], social scientists have learned to cope with this: they work with simple and "tidy models tested on toy systems", and they use the implementation and testing of these models "to capture important aspects of actual social systems". Agent-oriented software engineers should proceed in the same way. They should work with tidy and simple models tested on toy systems in order to construct systems with certain aspects and properties. The complexity should be in the result of the simulation, not in the model.

Social scientists need to create models of self-organizing MASs in order to understand real social systems with similar properties. Agent-oriented software engineers need to create models of self-organizing MASs in order to engineer systems with similar properties. Both face the same problems and can use similar tools. Agent-oriented software engineers are different from traditional software engineers, who try to prevent unpredictable behavior. They are as Scott Moss says [24] more like social simulators: "The unforeseen behavior that the engineer is trying to prevent is what the social simulator is interested in".

Thus the methods used in a Multi-Agent based Simulation (MABS) are the same we need to engineer MAS with emergent properties. Experiments, simulation and modeling are essential in order to design a self-organizing system and to engineer complex adaptive systems with emergent properties. Yet these simulations must always be guided and refined by the problem one wants to solve. For the social scientist, this is the phenomenon to be explained, for the AOSE engineer this is the aspect of the system to be realized (e.g. a "self-optimizing" or a "self-healing" system).

## 4. CONCLUSION

In this paper we examine the question if a general approach for the design and engineering of distributed self-organizing systems is possible. Can we really combine autonomy and self-organization on the one hand with engineering and design on the other hand? On first sight, it seems hardly possible to unite an unclear or changing purpose with an imposed purpose, or to combine self-organization with imposed organization.

We have argued that the design and engineering of self-organizing Multi-Agent Systems (MAS) is in fact difficult, and that this difficulty is not accidental or incidental, but rather an intrinsic and inherent problem related to MAS and autonomous agents in general. Engineering and autonomy interfere with each other, and other factors like "emergence" make the control of such systems difficult.

Yet the engineering of self-organizing systems or even complete "worlds" – systems that are so complex that they are an artificial world of their own – is indeed possible, if we apply the scientific method to the domain of engineering. The solution is therefore an *intelligent design* based on the classic *scientific method.* The problem of "engineering emergence" - to find a simple rule for a complex pattern – equals the problem of science in general: to explain complexity by describing complex natural phenomena with a minimum of primary principles, laws and rules. The central questions of the study of "emergence" and science are similar - how can you find simple local rules to generate specific higher levels of global organization - and therefore the answers or solutions are similar, too.

In concrete cases, the scientific method for the engineer can be an iterative goal-directed simulation where the goals are determined by high-level objectives and overall requirements, a two-phased approach based on extensive simulations which combines top-down analysis and bottom-up synthesis, hypothesis making with experimental verification, or simply theory with experimentation. One can invent many different names for the method, for example interactive man-machine method, iterative two-way approach, synthetic microanalysis, goal-directed simulation, stepwise refinement of educated guesses,… Important is the iterative, stepwise refinement of the model, the extensive use of simulations and experiments to verify hypotheses and models, and the constant combination of bottom-up synthesis with top-down analysis.

The AOSE engineer in agent-oriented software engineering can learn from the field of agent based modeling in the social sciences to construct small and tidy models which are testable on toy systems or in simulations. Simulations are essential for the verification of a system with unpredictable emergent properties, but not enough to find the right way through the jungle of complexity generated by recombination and composition on a large scale. The macroscopic view is also necessary to delineate possible states and configurations, to identify composite sub-systems on medium and large scales, and to set goals for microscopic simulations. The hard part is of course to find the right model at all. Depending on the global objectives and goal, an analysis of roles and sub-goals may not be enough. The review and recombination of already existing models and solutions can be helpful here.

This general approach is suitable in principle to create self-organizing systems and general distributed with self-* properties. Yet it is an open question what class of problems you can solve exactly, since self-organization has its limitations, too. Natural and social systems are not helpful here. Most of the really complex systems in nature are the result of pervasive evolution, which makes it difficult to distinguish between the effects of evolution and the effects of self-organization. Whenever we consider really complex natural forms, structures and patterns, we hit on traces of evolutionary influences, since nearly every complex life-form is subject to evolution. Therefore evolutionary algorithms offer a possible alternative solution, if the above-named methods reach their limits.